\documentclass[sigconf]{acmart}

\usepackage{graphicx}
\usepackage{subcaption}
\usepackage{enumitem}
\usepackage{hyperref}

\newcommand{\rev}[1]{#1}

\copyrightyear{2025}
\acmYear{2025}
\setcopyright{cc}
\setcctype{by}
\acmConference[VRST '25]{31st ACM Symposium on Virtual Reality Software and Technology}{November 12--14, 2025}{Montreal, QC, Canada}
\acmBooktitle{31st ACM Symposium on Virtual Reality Software and Technology (VRST '25), November 12--14, 2025, Montreal, QC, Canada}
\acmDOI{10.1145/3756884.3765987}
\acmISBN{979-8-4007-2118-2/2025/11}

\begin{document}
\title{AR-TMT: Investigating the Impact of Distraction Types on Attention and Behavior in AR-based Trail Making Test}

\author{Sihun Baek}
\affiliation{%
  % \department{Electrical and Computer Engineering}
  \institution{Duke University}
  \city{Durham}
  \state{NC}
  \country{USA}}
\email{sihun.baek@duke.edu}

\author{Zhehan Qu}
\affiliation{%
  % \department{Computer Science}
  \institution{Duke University}
  \city{Durham}
  \state{NC}
  \country{USA}}
\email{zhehan.qu@duke.edu}

\author{Maria Gorlatova}
\affiliation{%
  % \department{Electrical and Computer Engineering}
  \institution{Duke University}
  \city{Durham}
  \state{NC}
  \country{USA}}
\email{maria.gorlatova@duke.edu}
\renewcommand{\shortauthors}{Sihun, et al.}

\begin{teaserfigure}
    \centering
    \includegraphics[width=1\linewidth]{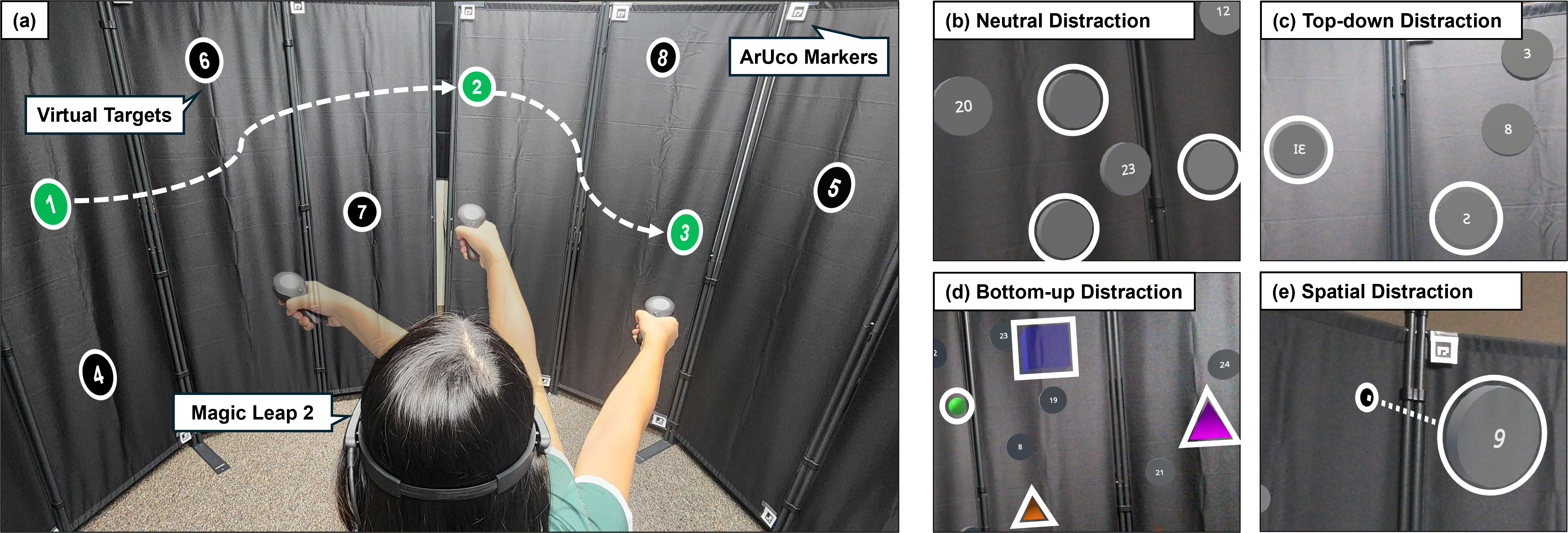} 
    \caption{
        The AR-TMT app. (a) App overview with numbers placed in circles laid on panels; (b–e) Example implementations of Neutral (ND), Top-down (TD), Bottom-up (BD), and Spatial Distractions (SD).
    }
    \label{fig:teaserfigure}
\end{teaserfigure}

\begin{abstract}
Despite the growing use of AR in safety-critical domains, the field lacks a systematic understanding of how different types of distraction affect user behavior in AR environments. To address this gap, we present \textbf{AR-TMT}, an AR adaptation of the Trail Making Test that spatially renders targets for sequential selection on the Magic Leap 2. We implemented distractions in three categories: top-down, bottom-up, and spatial distraction based on Wolfe's Guided Search model, and captured performance, gaze, motor behavior, and subjective load measures to analyze user attention and behavior. A user study with 34 participants revealed that top-down distraction degraded performance through semantic interference, while bottom-up distraction disrupted initial attentional engagement. Spatial distraction destabilized gaze behavior, leading to more scattered and less structured visual scanning patterns. We also found that performance was correlated with attention control (\( R^2 = .20\text{--}.35 \)) under object-based distraction conditions, where distractors possessed task-relevant features. The study offers insights into distraction mechanisms and their impact on users, providing opportunities for generalization to ecologically relevant AR tasks while underscoring the need to address the unique demands of AR environments.

\end{abstract}
% \vspace{-2cm}



\maketitle

\section{Introduction}

Understanding attention and distraction in AR is critical as users interact with visually complex scenes blending real and virtual elements, which can elevate the risk of distraction and potentially reduce task efficiency~\cite{dixon2013surgeons, kim2022assessing,wang2022inattentional}. Brief lapses in attention can have critical consequences in safety-critical domains such as healthcare and defense, where AR is increasingly used~\cite{biocca2007attention, livingston2011military}. To mitigate these risks and support effective performance, AR systems have introduced techniques such as diminished reality~\cite{herling2012pixmix, mori2017survey}, and visual guidance (e.g., highlights, directional cues, or spatial overlays)~\cite{warden2022visual, rusch2013directing}. Therefore, characterizing the mechanisms and effects of distraction in AR is essential for designing AR systems that enhance user focus and mitigate distraction.

However, many of these design strategies assume that certain stimuli are inherently distracting, often overlooking the specific visual features or contextual conditions that actually induce distraction in immersive environments~\cite{kim2022assessing,lee2025diminishar,tao2022integrating}. Additionally, previous works that incorporate attention theory remain application-driven and typically examine single factors in isolation, with limited consideration of how distraction mechanisms may differ in their effects~\cite{olk2018measuring, yang2019influences,ragan2015effects,biermeier2024measuring}. As a result, the field lacks a unified understanding of how these mechanisms contribute to attentional failures in AR. A related open question is how individual differences in attention control influence susceptibility to distraction in AR environments. While attention control has been extensively studied in traditional 2D tasks~\cite{burgoyne2023nature, unsworth2024individual, oberauer2024meaning}, the role of attention control in AR remains underexplored. Addressing these gaps is essential to identify general principles of distraction that can contribute to distraction-aware AR applications as well as a broader understanding of attention mechanisms in AR.

Therefore, we address the following research questions: 
\begin{itemize}[itemsep=0pt, topsep=0.3pt, parsep=1pt, leftmargin=0.15in]
    \item \textbf{RQ1:} How can distraction in AR be categorized by attentional mechanisms?
    \item \textbf{RQ2:} How do different distractions affect user attention and behavior as measured with the AR headset?
    \item \textbf{RQ3:} How do individual differences in attention control influence vulnerability to distraction in AR?
\end{itemize}

To investigate these research questions, we developed AR-TMT, an AR adaptation of the Trail Making Test~\cite{bowie2006administration, mitrushina2005handbook,reitan1958validity} that systematically manipulates distraction types based on the Guided Search model~\cite{wolfe2017five, wolfe2021guided, wolfe2020visual} on the Magic Leap 2. We operationalized distraction through distinct attention-guiding factors as top-down, bottom-up, and scene-based guidance (RQ1), and examined their effects on user behavior across performance, visual behavior, motor behavior, and subjective measures (RQ2). To assess attention control capacity, we incorporated the Flanker Squared Test~\cite{burgoyne2023nature}, a validated psychometric measure of attention control, and examined how these scores relate to variation in AR-TMT performance and behavior (RQ3). We evaluated AR-TMT in a user study with 34 participants to examine these research questions.

This paper makes three main contributions: (1) We instantiated Guided Search–based distraction mechanisms by adapting the TMT to AR within a unified framework, enabling systematic manipulations of top-down, bottom-up, and spatial distractions. (2) We showed that different types of distraction produce distinct behavioral patterns across performance, gaze, motor behavior, and subjective ratings, as measured through AR-TMT. (3) We found that performance under object-based distraction conditions was correlated with Attention Control Score (ACS) (\( R^2 = .20 \)–.35), highlighting that attentional traits are associated with susceptibility to distraction from task-irrelevant objects in AR.

The remainder of the paper is organized as follows. Section~\ref{sec:related} reviews related work on distraction in XR, the Trail Making Test (TMT), and attention control. Section~\ref{sec:approach} details the design of the AR-TMT task, experimental framework, and hypotheses. Sections~\ref{sec:userstudy} and \ref{sec:result} describe the study setup and results, respectively. Sections~\ref{sec:discussion} and \ref{sec:limF} discuss key findings, implications, limitations, and future work. Finally, Section~\ref{sec:conclusion} summarizes our contributions and concludes the paper.

\section{Related Work}
\label{sec:related}
\paragraph{Distraction in XR}
Many prior studies have focused on understanding the nature and impact of distraction in XR environments. 
Studies have examined the effect of background distractions on task performance and visual fatigue~\cite{arefin2021effects}, and the use of distraction to reduce pain ~\cite{malloy2010effectiveness, glennon2018use}. In addition, the influence of distracting AR elements on emergency preparedness and puzzle solving~\cite{mirza2024exploring, qu2024looking}, and the potential distraction during driving from AR head-up displays (HUDs)~\cite{kim2022assessing} has been investigated. 
In VR, prior work has compared distractions in educational settings~\cite{fisher2024comparing}, and examined internal and external distraction during learning using distraction levels and multimodal features (e.g., eye tracking, EEG)~\cite{asish2022detecting, asish2024classification}.
Similarly, Olk et al.~\cite{olk2018measuring} showed that increasing target–distractor similarity and incongruent flankers impair visual search performance in VR. Interestingly, the findings suggest that distraction does not always affect performance~\cite{arefin2021effects, qu2024looking}, and the specific characteristics that make distractions consequential in AR remain underexplored.

\paragraph{Trail Making Test and Attention Control} The Trail Making Test (TMT) is a \rev{neuropsychological tool} to assess cognitive flexibility and sequencing~\cite{sanchez2009construct, jacobson2011fmri, bowie2006administration}. To improve ecological validity, recent studies have adapted the TMT to immersive environments. A 3D supermarket version has demonstrated strong convergence with the classic TMT while enhancing realism~\cite{malegiannaki2021can}, and VR-based implementations have improved usability and engagement, particularly for clinical populations~\cite{giatzoglou2024trail, gounari2025trail}. In parallel, research has established reliable tools for quantifying attention control~\cite{burgoyne2023nature, lee2025precise, oberauer2024meaning, mashburn2020individual}, and prior work has shown that attention control moderates the impact of distraction on academic performance~\cite{deepa2022moderating}. Building on these efforts, we use the TMT as a structured framework to examine how different types of distraction affect user behavior and whether sensitivity to distraction types is associated with individual differences in attention control.

\section{Approach}
\label{sec:approach}
\begin{figure}[t]
  \centering
  \includegraphics[width=0.9\linewidth]{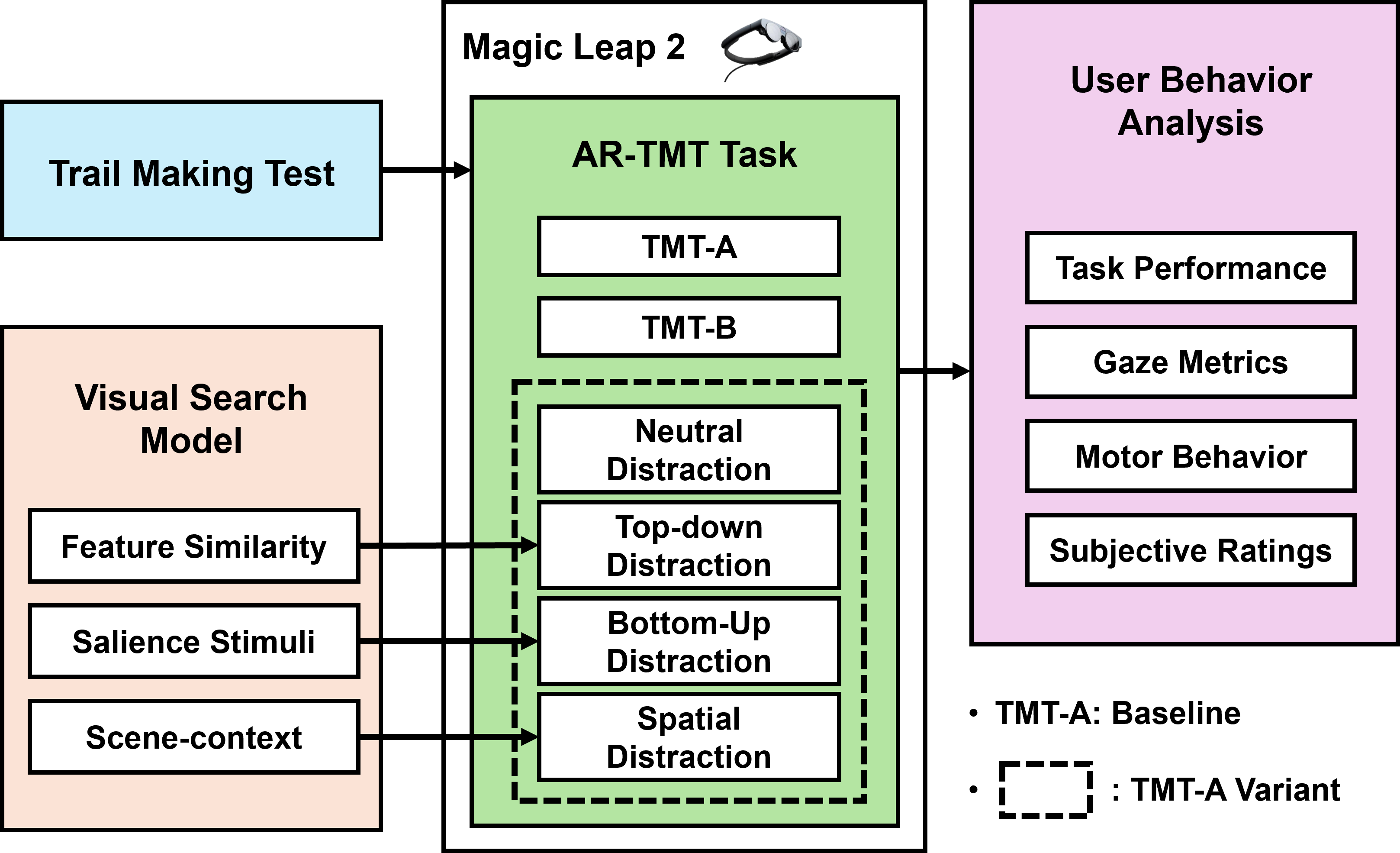}
  \caption{Overview of our experimental framework.
}
  \label{fig:system}
\end{figure}

This section outlines the theoretical and methodological foundations of our study. We first describe the adaptation of the Trail Making Test into an augmented reality task (AR-TMT), followed by the design of distraction stages grounded in the Guided Search model~\cite{wolfe2017five, wolfe2021guided, wolfe2020visual}. Figure~\ref{fig:system} provides an overview of the experimental framework, including task structure and distraction conditions.

\subsection{AR-TMT Task Adaptation} TMT is a popular neuropsychological tool used both independently for screening neurological and cognitive impairments such as dementia, brain injury, or ADHD, and as a component of comprehensive neuropsychological test batteries~\cite{bowie2006administration, mitrushina2005handbook,reitan1958validity, tombaugh2004trail}. The test assesses cognitive domains such as visuomotor speed, working memory, task-switching, and visual attention, and is traditionally administered using paper or a computer monitor.

The test consists of two parts: \textbf{TMT-A} and \textbf{TMT-B}. In TMT-A, individuals are instructed to connect 25 numbered targets (1–2–3...) in ascending order as quickly and accurately as possible. TMT-B increases complexity by requiring participants to alternate between numbers and letters in alphanumeric order (1–A–2–B–3–C...), placing greater demands on working memory, task switching, and cognitive flexibility~\cite{sanchez2009construct, salthouse2011cognitive, jacobson2011fmri}. The primary performance metric is completion time, including time spent correcting errors. While TMT-A mainly reflects visuoperceptual abilities, TMT-B provides a more sensitive index of executive control~\cite{sanchez2009construct,salthouse2011cognitive}.

The TMT is a well-suited task paradigm to examine distraction mechanisms, given its structured sequence that engages multiple cognitive processes and has the potential to generalize to ecologically relevant AR tasks. Unlike the simplified visual search tasks where users typically locate a single target among distractors, the TMT requires continuous updating of working memory, monitoring of sequential order, and goal-directed physical responses~\cite{sanchez2009construct,salthouse2011cognitive}. These demands closely resemble tasks in AR applications, such as navigating hierarchical menus, activating spatially distributed virtual controls, or following step-by-step assembly instructions. Additionally, recent VR adaptations of the TMT have preserved its diagnostic sensitivity while increasing ecological validity, further supporting its suitability for AR contexts~\cite{malegiannaki2021can, giatzoglou2024trail, plotnik2021multimodal}.

To extend the capabilities of the TMT into AR, we adapted both TMT-A and TMT-B for an AR environment. Our implementation preserves the essential structure of the classic test while introducing controlled distraction stages based on the Guided Search model. Within AR-TMT, TMT-A serves as a baseline due to its relatively low cognitive demand, enabling clear comparisons with distraction stages. TMT-B serves as an additional condition to contextualize distraction-related effects against a well-established measure of cognitive flexibility. This design enables systematic manipulation of attentional interference while building on the task’s validated diagnostic foundation.

\subsection{Attentional Framework and Distraction Types in AR-TMT}
To investigate how different types of distraction impact attention and behavior in AR environments, we draw from Wolfe’s Guided Search model~\cite{wolfe2017five, wolfe2021guided, wolfe2020visual}, a framework that explains visual search as a process guided by multiple attention-capturing factors: bottom-up salience, top-down feature guidance, scene-based context, search history (priming), and reward. While these factors are traditionally discussed as guidance mechanisms, we examined them through the lens of distraction. In AR settings, where virtual content competes with real-world stimuli, these same mechanisms can become sources of interference that disrupt task-relevant attention.

In this study, we focus on three factors in the framework: \textbf{top-down feature guidance}, \textbf{bottom-up salience}, and \textbf{scene-based context}. 
% top-down features similairty
Top-down (goal-driven) feature similarity can cause semantic interference when distractors share attributes with the target.
% bottom-up salience
Bottom-up salience can divert attention via highly distinctive but task-irrelevant visual stimuli.
% scene-based context
Violations of scene-based context (spatial or semantic) can disrupt guidance and hinder search efficiency.
% Why they are chosen
Top-down feature guidance and bottom-up salience were selected as core mechanisms of attention with strong theoretical grounding in visual search (e.g., salient AR pop-ups or task-relevant icons competing with targets). Scene-based context was included because AR relies on 3D spatial organization and environmental anchoring, making spatial disruptions (e.g., misaligned spatial overlays or disappearing anchors) more critical than traditional 2D tasks or static 3D environments. In contrast, search history and reward-based guidance were not considered because their inclusion would overload the task design and complicate experimental control in a within-subjects design. By framing attention-guiding factors as structured distraction types, AR-TMT provides a controlled framework to examine their influence on user behavior and yields insights that generalize to ecologically relevant AR tasks.

\subsection{\rev{Hypotheses}}
\label{sec:hypotheses}
We formulated the following hypotheses based on the framework.

\textbf{H1.} \textit{Top-down distraction} draws attention through semantic interference, inducing performance degradation.

\textbf{H2.} \textit{Bottom-up distraction} captures attention transiently at onset, with its effect diminishing thereafter.

\textbf{H3.} \textit{Spatial distraction} interferes with scene guidance, reducing search efficiency and randomizing gaze patterns.

\textbf{H4.} \textit{Attention control} is associated with performance under top-down and bottom-up distractions.

\rev{Our hypotheses (H1–H3: distraction mechanisms, RQ2; H4: attention control, RQ3) are tested across four measures. \textit{Performance measures} (completion time, reaction time) capture task efficiency, and \textit{visual behavior measures} (fixation metrics, saccade metrics, and gaze entropy) reflect visual attention and search organization. \textit{Motor behavior measures} (controller speed, movement entropy) indicate motor engagement under cognitive load, and \textit{subjective measures} (self-reported ratings) assess perceived cognitive load and distraction. }

\rev{Top-down distraction (H1) is expected to hold visual attention on distractors with effects on visual behavior (e.g., longer fixations, reduced saccade velocity), leading to longer completion time and higher perceptual load due to semantic interference~\cite{olk2018measuring, horstmann2019dwelling}. Bottom-up distraction (H2) is expected to capture attention transiently at onset, affecting early-stage performance (e.g., reaction time for the first target), but have weaker overall effects since salient distractors can be suppressed afterward~\cite{gaspelin2015direct, chen2023saliency}. Spatial distraction (H3) is expected to reduce search efficiency by disrupting scene context, with effects on gaze distribution (e.g., more randomized gaze patterns), leading to longer completion time and reduced comfort with target placement~\cite{sisk2019mechanisms, poupyrev20003d}. Attention control (H4) is expected to be associated with performance under top-down and bottom-up distraction, as these forms of distraction are conceptually aligned with the construct of attention control~\cite{oberauer2024meaning, burgoyne2023nature}.}

\section{User Study}
\label{sec:userstudy}

This section details the AR-TMT user study, including the experimental setup, task implementation, and study procedure to empirically evaluate our research questions and hypotheses.

\subsection{Experimental Setup}
\paragraph{Hardware and Software.} We deployed AR-TMT on a Magic Leap 2 AR headset~\cite{curtis2022unveiling}, while a laptop with a 13th-Gen Intel Core i5-1335U processor managed the experiment and controlled the application remotely. Device management and data transfer were handled via Magic Leap Hub 3, a developer tool to interact with the Magic Leap system. We developed AR-TMT in Unity (2022.3.48f1) using the Magic Leap OpenXR Unity SDK. We implemented three AR interfaces: (i) a main menu for stage selection and rest periods, (ii) an instruction screen for task guidance, and (iii) a survey to collect subjective ratings without headset removal. 
% Data Collection
% Performance
For data collection, we recorded trial completion times and the sequence of target selections (timestamp, hit/miss outcome, and selection order) to capture task performance.  
% Eye tracking/controller/head
We also recorded eye tracking, controller motion, and head motion at 60 Hz. Fixations were identified using the I-VT algorithm~\cite{salvucci2000identifying} with a 30°/s velocity threshold, while faster movements were classified as saccades and blinks were excluded.
% subjective response 
Through the AR interface, participants reported subjective ratings of perceived distraction from virtual stimuli and the physical environment, mental demand, and comfort with target placement.

\paragraph{Physical Setup.} % Room
The user study was conducted in a dedicated room without sunlight to minimize environmental variability.
% Walls 
To simulate real-world spatial constraints, we installed \textit{two divider structures}, each composed of four adjustable panels, arranged in a fan-shaped, semi-enclosed 180° arc, serving as tangible spatial references (Fig.~\ref{fig:teaserfigure}). Each panel measured 170~$\times$~52~cm and was equipped with two 3~$\times$~3~cm ArUco markers (placed at the top-left and bottom-right corners of each panel) for 3D tracking. Virtual targets were anchored to this wide horizontal field to encourage natural head and eye movements. The dividers had minimal visual detail (plain black surfaces) to avoid introducing unintended visual distractions or semantic cues. Compared to VR-based TMT studies~\cite{giatzoglou2024trail, gounari2025trail, plotnik2021multimodal}, we designed our setup to preserve real-world spatial semantics and structure by anchoring virtual objects to physically grounded surfaces. 
% Chair
We placed a \textit{rolling swivel chair} centrally in front of the panels to support comfortable 6-DoF interaction. Sitting ensured natural eye-level alignment with targets while allowing head and body rotation within a 180° horizontal arc and ~40–45° vertically. 
% As the task primarily relied on rotational scanning, positional movement was minimal. 

\subsection{AR-TMT Task Implementation}
\label{sec:implementation}
\paragraph{Overview of Stages} We implemented six stages for distraction analysis in the AR-TMT app: two adapted from the classic TMT, followed by four distraction conditions (neutral, top-down, bottom-up, and spatial distraction) based on the TMT-A format. We included a neutral distraction condition to isolate the effect of visual presence without goal-driven interference or high saliency. To ensure experimental control, all targets and distractors were rendered as virtual objects, with targets implemented as plain light-gray discs (10 cm diameter, 2 cm thick). A total of 25 numbered targets were randomly positioned on the panels, with a minimum spacing of 1.5 times the targets' diameter to reduce visual clutter and prevent overlap.

\paragraph{Target Arrangement} To ensure that performance differences primarily reflect distraction effects rather than target arrangement, we varied target arrangements across stages while maintaining comparable visuospatial search demands. Targets were randomly distributed across all eight panels, subject to two constraints: (1) to minimize order effects, consecutive targets were not placed on the same panel, preventing localized sequences that could lower spatial demand; and (2) to minimize placement effects, targets were evenly distributed across panels (2–4 targets per panel), reducing clustering that might bias difficulty. To verify comparability across stages, we measured angular head movement between consecutive targets in each participant’s selection sequence and then averaged these values across participants. Variability across stages was low (mean = 53.78$^\circ$, SD = 4.15$^\circ$, CV $\approx$ 7.71\%), indicating comparable visuospatial demands across stages. Further analysis of arrangement effects is provided in Section~\ref{sec:perofrmanceanalysis}.

\paragraph{Distraction Conditions} Each distraction condition was implemented as a separate stage as shown in Fig.~\ref{fig:teaserfigure}. \textit{Neutral Distraction (ND) stage} included 20 additional objects identical in size and color to the targets but without numbers, allowing us to assess the impact of visual presence without goal-driven interference or high saliency. \textit{Top-down Distraction (TD) stage} included 20 semantically interfering distractors resembling target characters but incorrect (lookalikes), such as mirrored “S”, stacked “I” and “O” (e.g., IO, II), and rotated elements (e.g., $\infty$, $\backsim$). \textit{Bottom-up Distraction (BD) stage} introduced 20 highly salient distractors varying in shape (triangle, square, sphere), size (0.5--1.5$\times$ target size), and color (eight distinct hues) with one third of them dynamic (oscillating, flickering, or spinning) to enhance bottom-up attentional capture. \textit{Spatial Distraction (SD) stage} removed wall-based spatial cues by placing targets in mid-air rather than anchoring them to the panels. Targets were distributed within a 140$^\circ$ cone in front of the user (0.8–3 m depth; $\pm$0.7 m vertical with a 0.2 m downward bias) to match the directional search space of other stages. To prevent overlap, we enforced a minimum separation of 6$^\circ$ (angular) and 0.4 m (spatial).\footnote{AR-TMT app demonstration video: \url{https://github.com/Duke-I3T-Lab/AR-TMT.git}}

\subsection{Study Procedure}

\rev{The study consisted of a pre-test session and the AR-TMT task, with a survey given after each stage of AR-TMT.} The study protocol was approved by Duke University's Institutional Review Board.

\paragraph{Pre-Test Session} Upon arrival, participants provided informed consent, completed a demographic questionnaire (age and gender), and received a verbal explanation of the study. 
% Attention Control
Attention control was assessed using two established measures. The primary measure was the \textit{Flanker Squared Test}~\cite{burgoyne2023nature}, a validated behavioral task strongly associated with a latent attention control factor. As a supplementary measure, participants completed the \textit{Attention Control Scale (ATTC)}~\cite{derryberry2002anxiety}, a 20-item self-report questionnaire evaluating attentional focusing and shifting on a four-point Likert scale.
% Conventional TMT
To compare AR-TMT with the conventional task, participants first completed a monitor-based 2D version of TMT that we implemented to provide standardized administration, automated scoring, and practice trials for familiarization.
% visuomotor speed task
Participants then completed headset calibration for head and eye tracking and performed a visuomotor reaction task, adapted from the Serial Reaction Time Task~\cite{robertson2007serial,heyes2002motor}, to assess reaction speed in AR. In each of 30 trials, a single unnumbered target appeared, and once selected, a new target appeared at a random visible location. Reaction times were recorded for each trial. Motor Skill Score (MSS) was defined as the inverse of mean reaction time, with higher values indicating better performance. Visual targets were identical to those in the main task, providing implicit training with the AR-TMT interface.

\paragraph{AR-TMT Task.} Due to the headset’s limited field of view (FOV), participants were explicitly instructed to turn their heads to locate peripheral targets. Before beginning, participants scanned eight ArUco markers mounted on surrounding divider panels to establish the 3D spatial layout for anchoring virtual content. The session began with a stage selection interface presented in AR. After completing the visuomotor speed task, participants proceeded through the AR-TMT stages in an order determined by a Latin-square design. Each stage was preceded by an instruction interface describing the task objectives, interaction method, and movement guidelines. A 60-second break followed each stage to mitigate cognitive fatigue. After each stage, participants completed four subjective rating items on a 7-point Likert scale to assess (1) cognitive load, (2) distraction from virtual objects, (3) distraction from the physical environment, and (4) comfort with target placement (reverse-scored to reflect perceived spatial interference). The mental demand item was adapted from the NASA-TLX~\cite{hart1988development}, while the remaining items were designed to capture participants’ perception of distraction intensity.

\begin{figure}[tbp]
    \centering
    \begin{subfigure}{0.48\linewidth}
        \centering
        \includegraphics[width=\linewidth, trim=7pt 0pt 5pt 5pt, clip]{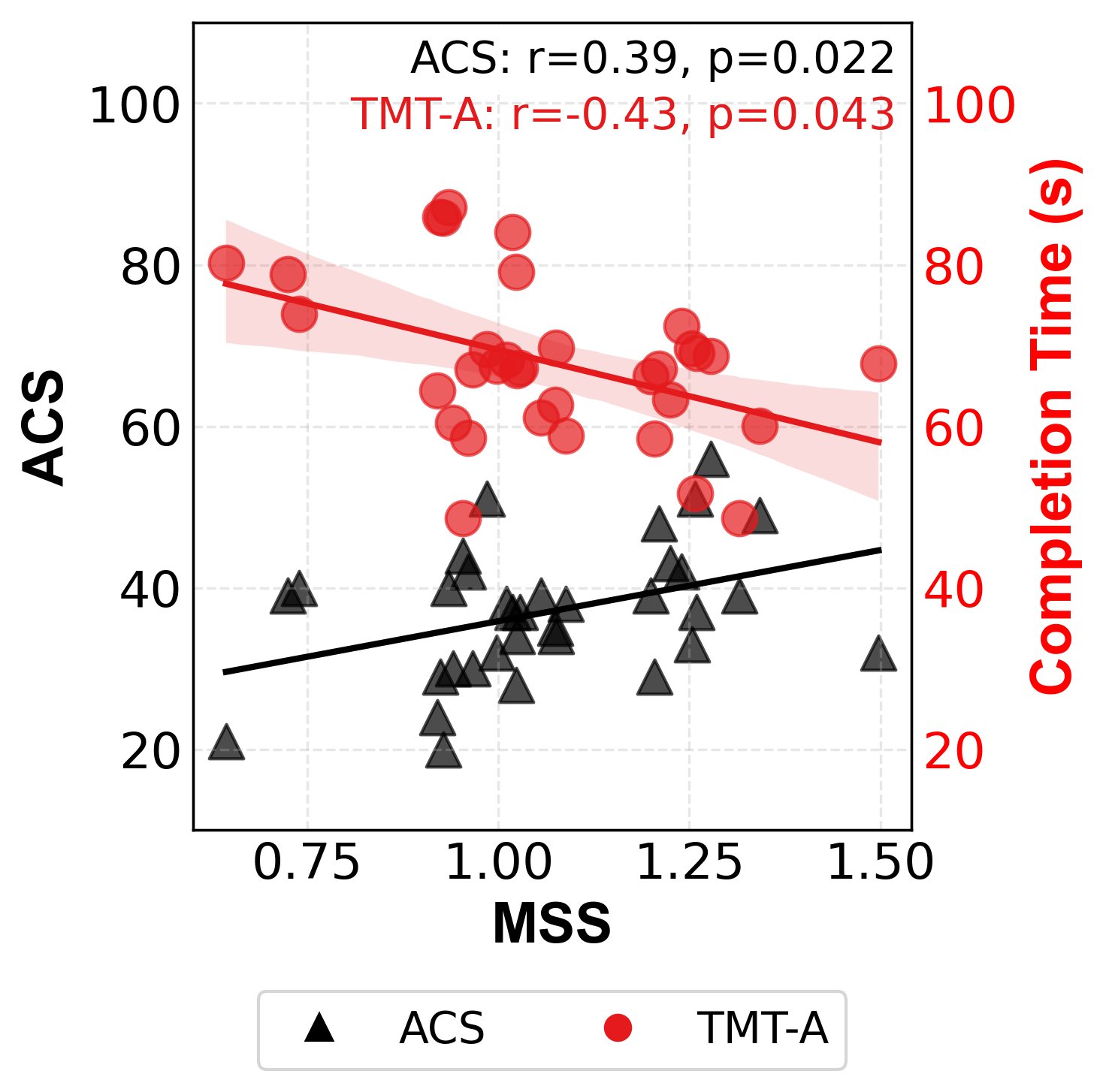}
        \caption{MSS vs. ACS and Stage A.}
        \label{fig:motrspeed}
    \end{subfigure}
    \hfill
    \begin{subfigure}{0.48\linewidth}
        \centering
        \includegraphics[width=\linewidth, trim=7pt 0pt 5pt 5pt, clip]{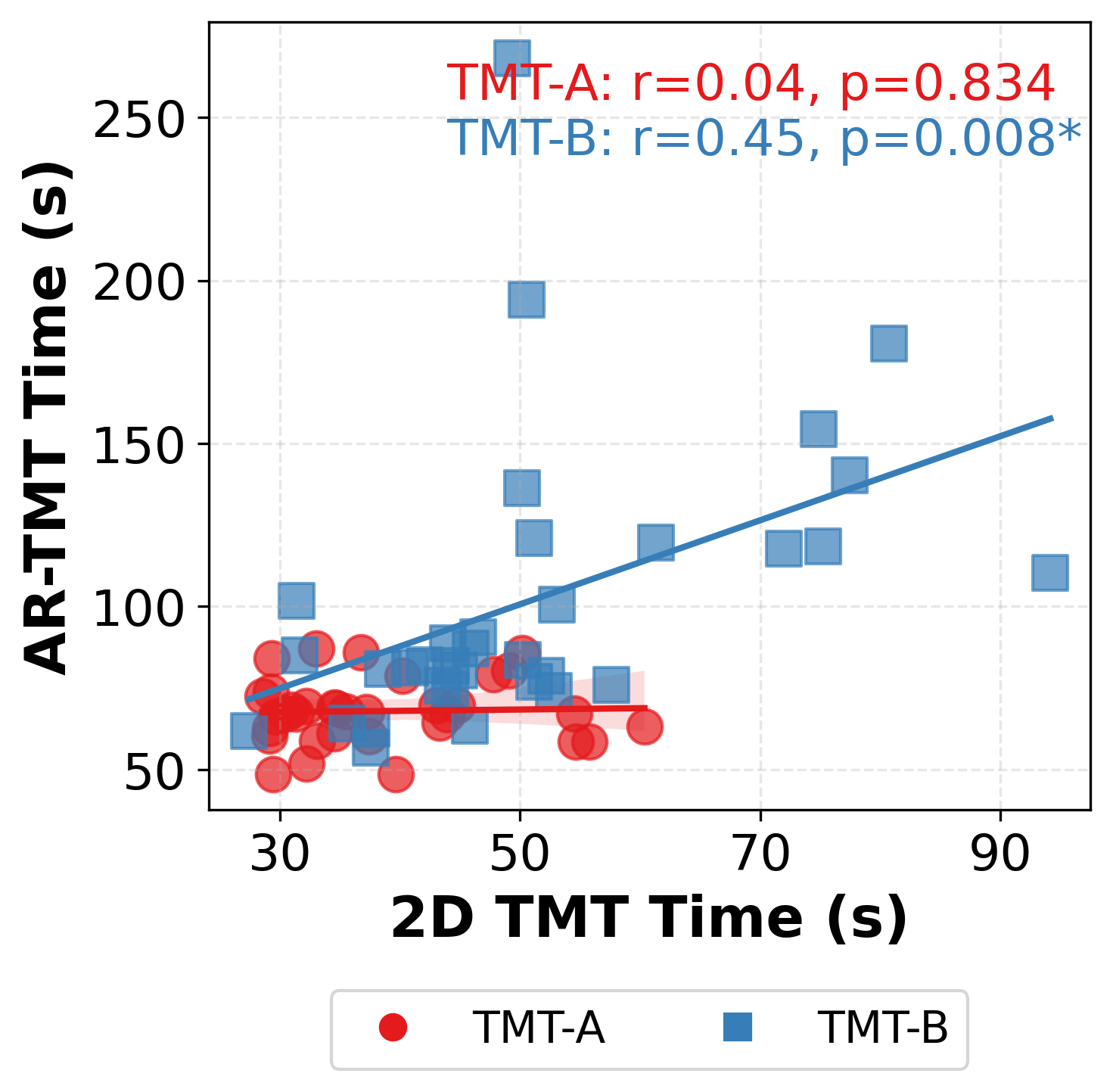}
        \caption{AR vs. 2D TMT.}
        \label{fig:classicTMT}
    \end{subfigure}
        \caption{ (a) Correlation analysis of Motor Skill Score (MSS) with Attention Control Score (ACS) and completion time of TMT-A in AR (Holm-corrected $p$ value for TMT-A). (b) Correlation analysis between AR-TMT and 2D TMT.
        }

    \label{fig:Motor&Classic TMT}
\end{figure}

\paragraph{Participants} We recruited 34 participants (mean \rev{age = $25 \pm 5$ years}, range: 18--44 years; 19 female) from the university community via email announcements, electronic flyers, and posters. Thirteen participants wore glasses, and four reported an ADHD diagnosis. Prior AR experience was minimal: 23 participants had never used optical see-through AR devices, 8 had tried them once or twice, and 4 had used them infrequently (less than once per week). The average ACS was $37.06 \pm 8.34$, and the average ATTC score was $54.82 \pm 6.72$. Notably, there was no significant correlation between the two measures ($p > .90$), suggesting a lack of correlation between ACS and ATTC. Additionally, ACS was positively correlated with MSS ($r = .39$, $p = .02$), as shown in Fig.~\ref{fig:motrspeed}, indicating that faster motor responses were associated with greater attention control. This aligns with prior work linking processing speed and attention~\cite{burgoyne2023nature}.

\subsection{Benchmarking Against 2D TMT}
AR-TMT introduces a different mode of interaction (e.g., head-mounted display and controller) compared to the 2D TMT, as previously explored in VR-based TMT studies~\cite{plotnik2021multimodal, gounari2025trail}. We compared AR-TMT with 2D TMT to contextualize user performance and examine whether the established functions of TMT-A (visuoperceptual ability) and TMT-B (cognitive flexibility) appear to generalize across modalities. This allows clearer interpretation of distraction effects in AR by anchoring them to known behavioral benchmarks.

As shown in Fig.~\ref{fig:classicTMT}, performance on TMT-A showed little to no correlation between the AR and 2D formats (\( r = 0.04 \), \( p = .83 \)), indicating limited cross-modal consistency for visuoperceptual abilities. In contrast, performance on TMT-B revealed a moderate correlation with 2D TMT-B ($r = 0.45$, $p = .008$), suggesting that cognitive flexibility demands may generalize more robustly across interaction modalities. While many prior studies reported correlations for both TMT-A and TMT-B across modalities~\cite{malegiannaki2021can, plotnik2021multimodal, gounari2025trail}, our results align with \cite{plotnik2021multimodal}, which found that only TMT-B showed cross-modal correlation in healthy young adults using head-mounted VR systems. The discrepancy in TMT-A performance may reflect differences in interface familiarity among younger users. The majority of young people are well-practiced in 2D coordination through monitor-based interactions, whereas they have far less experience with the 3D visuomotor interactions required in AR environments. Supporting this, MSS was correlated with TMT-A in AR ($r = .43$, $p = .043$), but not with 2D TMT-A (Fig.~\ref{fig:motrspeed}). These results highlight that while cognitive flexibility transfers across modalities, visuoperceptual abilities in AR are shaped by unique spatial and interaction demands, emphasizing the need to account for modality-specific factors when interpreting distraction-related behavior.

\section{Results}
\label{sec:result}

This section reports results across task performance, visual behavior, motor behavior, and subjective measures using mixed-effects models to examine distraction and attention control effects.

\subsection{Performance Analysis}
\label{sec:perofrmanceanalysis}
\paragraph{Completion Time} To investigate the impact of distraction conditions and whether individual differences in attention control are related to task performance (completion time), we fit a Linear Mixed-effects Model (LMM) with Stage, ACS, and their interaction as fixed effects, \rev{and random intercepts for participants to account for repeated measures within subjects.}  We also conducted stage-wise linear regression analyses to test in which stages ACS was associated with completion time.

\begin{figure}[tbp]
    \centering
    \begin{subfigure}[t]{0.48\linewidth}
        \centering
\includegraphics[width=\linewidth]{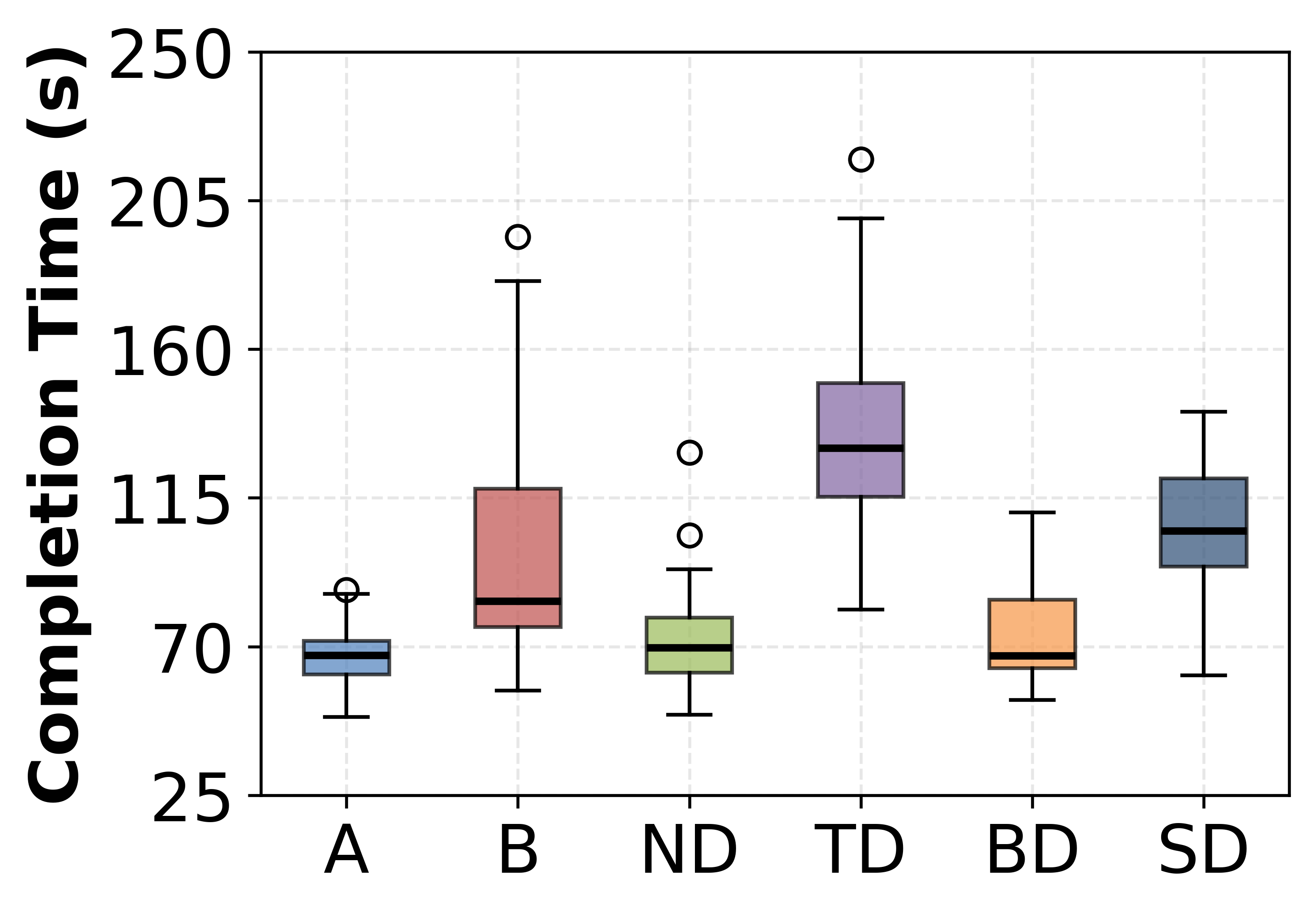}
        \caption{Completion time by stage.}
        \label{fig:completiontime}
    \end{subfigure}
    \hfill
    \begin{subfigure}[t]{0.48\linewidth}
        \centering
        \includegraphics[width=\linewidth]{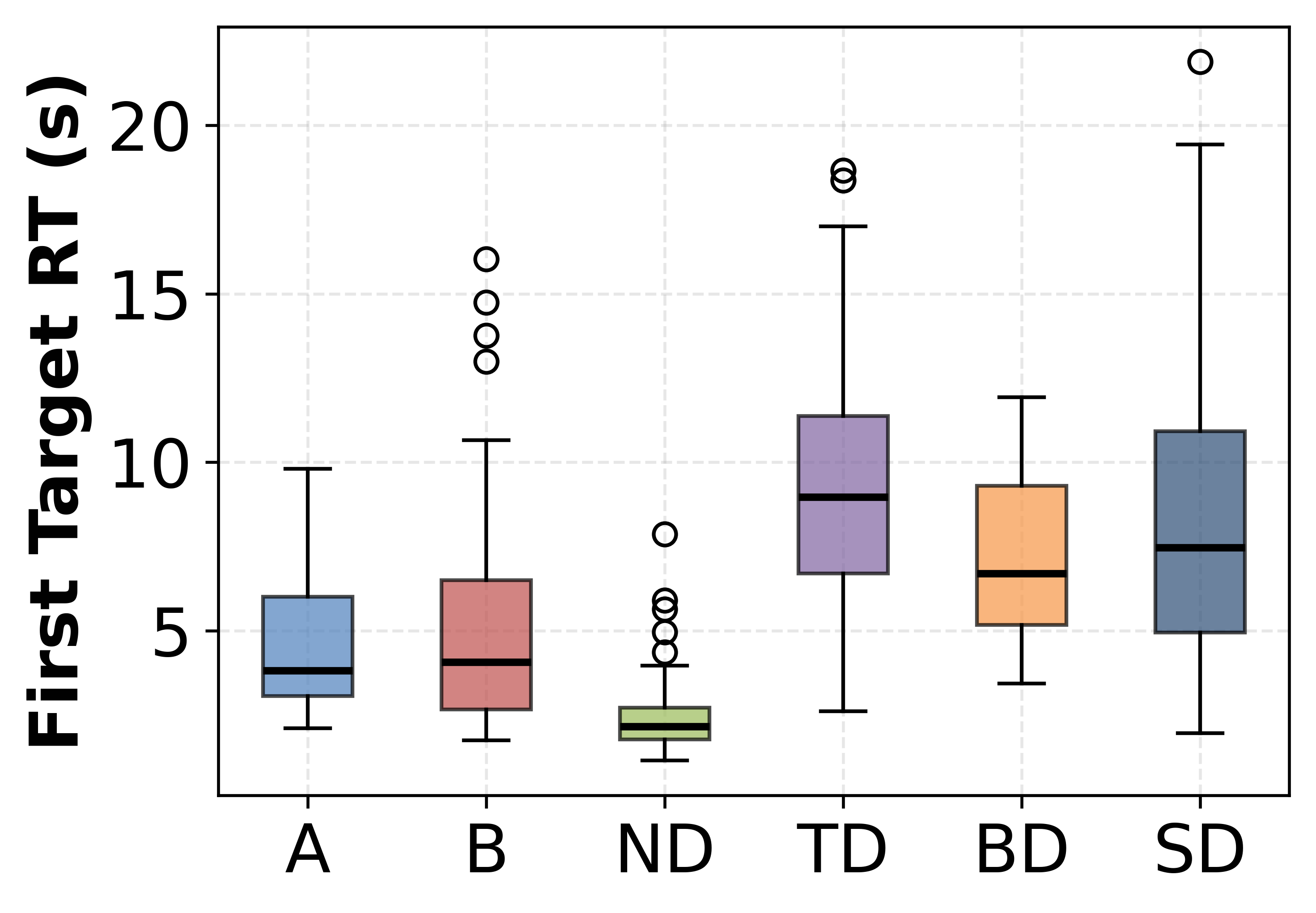}
        \caption{First target RT by stage.}
        \label{fig:firstRT}
    \end{subfigure}
    \caption{Boxplots of performance metrics (completion time and the first target RT) across distraction stages.}
    \label{fig:boxplots_stage}
\end{figure}

% description of meanvalue
Fig.~\ref{fig:completiontime} shows a boxplot of completion times across AR-TMT stages. Mean completion times ranged from \(67.92 \pm 9.90\)~s in \mbox{stage A} (baseline) to \(102.01 \pm 44.19\)~s in B, \(73.91 \pm 18.70\)~s in ND, \(135.36 \pm 29.77\)~s in TD, \(76.17 \pm 20.16\)~s in BD, and \(107.53 \pm 19.75\)~s in SD. 
% lmm result
The LMM results revealed longer completion times in B ($\beta = 58.96$, \rev{95\% CI [9.73, 108.19]}, $p = .020$) and TD stages ($\beta = 109.90$, \rev{95\% CI [60.67, 159.13]}, $p < .001$), relative to stage A. 
% posthoc analysis
Holm-corrected pairwise comparisons of estimated marginal means revealed that completion times were longer in TD compared to all other stages (\rev{$\Delta = 27.8$--$67.4$~s}, $p < .001$). SD also showed longer times than A, BD, and ND (\rev{$\Delta = 31.4$--$39.6$~s}, $p < .001$), while both BD and ND were faster than TD (\rev{$\Delta = -59.2$ to $-61.5)$~s}, $p < .001$). Additionally, SD was faster than TD (\rev{$\Delta = -27.8$~s}, $p < .001$).

\begin{figure}[tbp]
    \centering
    \includegraphics[ width=\linewidth, trim=0pt 10pt 0pt 0pt, clip]{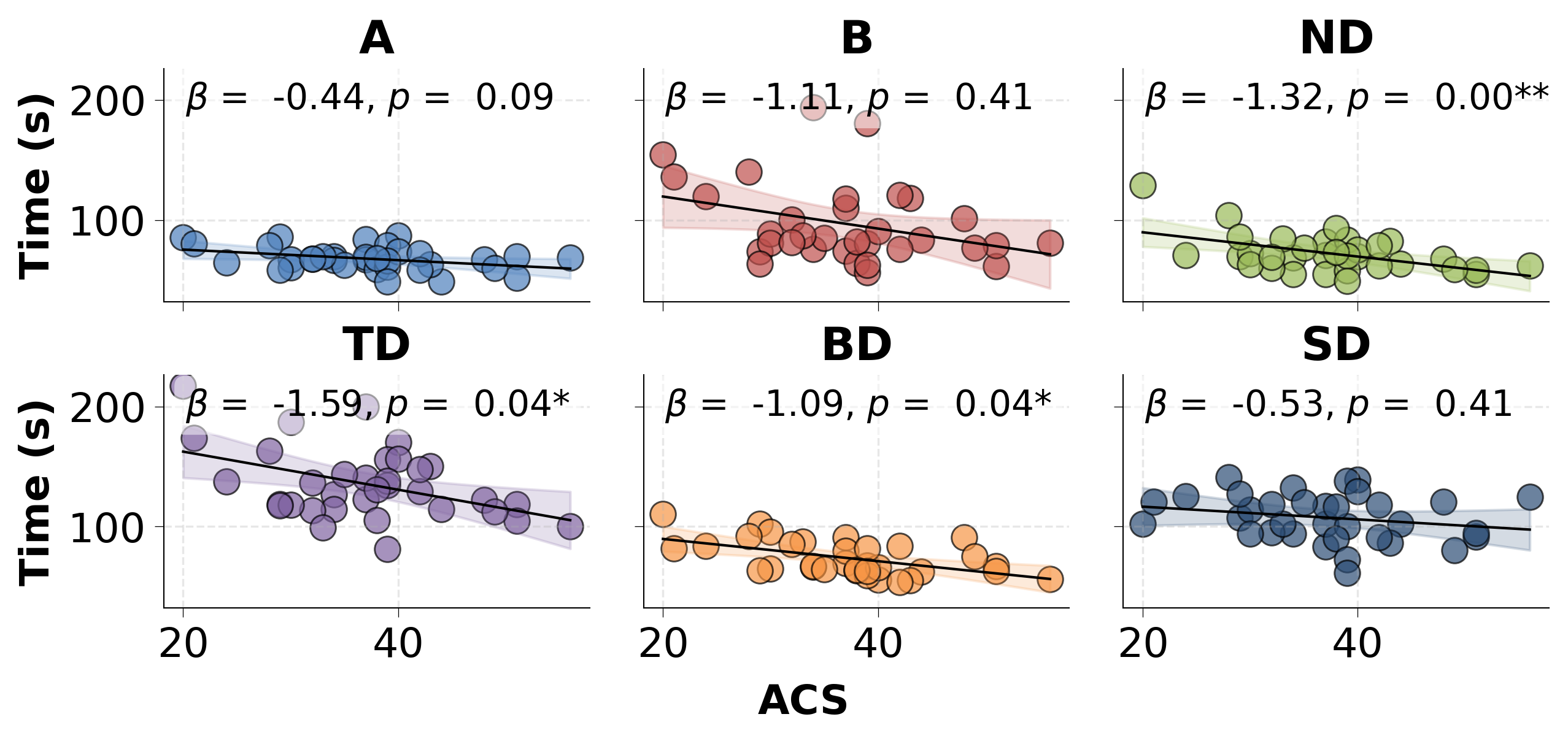}
    \caption{Linear regression between Attention Control Scores (ACS) and completion time across distraction stages with Holm-corrected $p$-values; significance is marked (* = $p < .05$, ** = $p < .01$).}
    \label{fig:completion_time_vsACS}
\end{figure}

Figure~\ref{fig:completion_time_vsACS} presents the stage-wise linear regression results, showing that Attention Control Score (ACS) was negatively associated with completion time in ND ($\beta = -1.32$, $p = .0015$, $R^2 = .35$), BD ($\beta = -1.09$, $p = .0357$, $R^2 = .21$), and TD ($\beta = -1.59$, $p = .0357$, $R^2 = .20$) after Holm correction, with no significant associations in A, B, or SD stages ($p > .09$). Stage B, emphasizing cognitive flexibility, showed cross-modal consistency from 2D TMT but was not correlated with ACS, whereas distraction stages (TD, ND, BD) were more sensitive to individual differences in attention control.

These findings suggest that top-down distraction had the greatest impact on performance, followed by spatial distraction, while ND and BD showed minimal differences from the baseline. Moreover, individuals with lower attention control were especially vulnerable under object-based distraction conditions (ND, BD, TD), whereas attention control was not predictive of performance in the baseline or spatially disorganized conditions.

\begin{figure}[tbp]
    \centering
    \includegraphics[width=\linewidth, trim=20pt 10pt 0pt 0pt, clip]{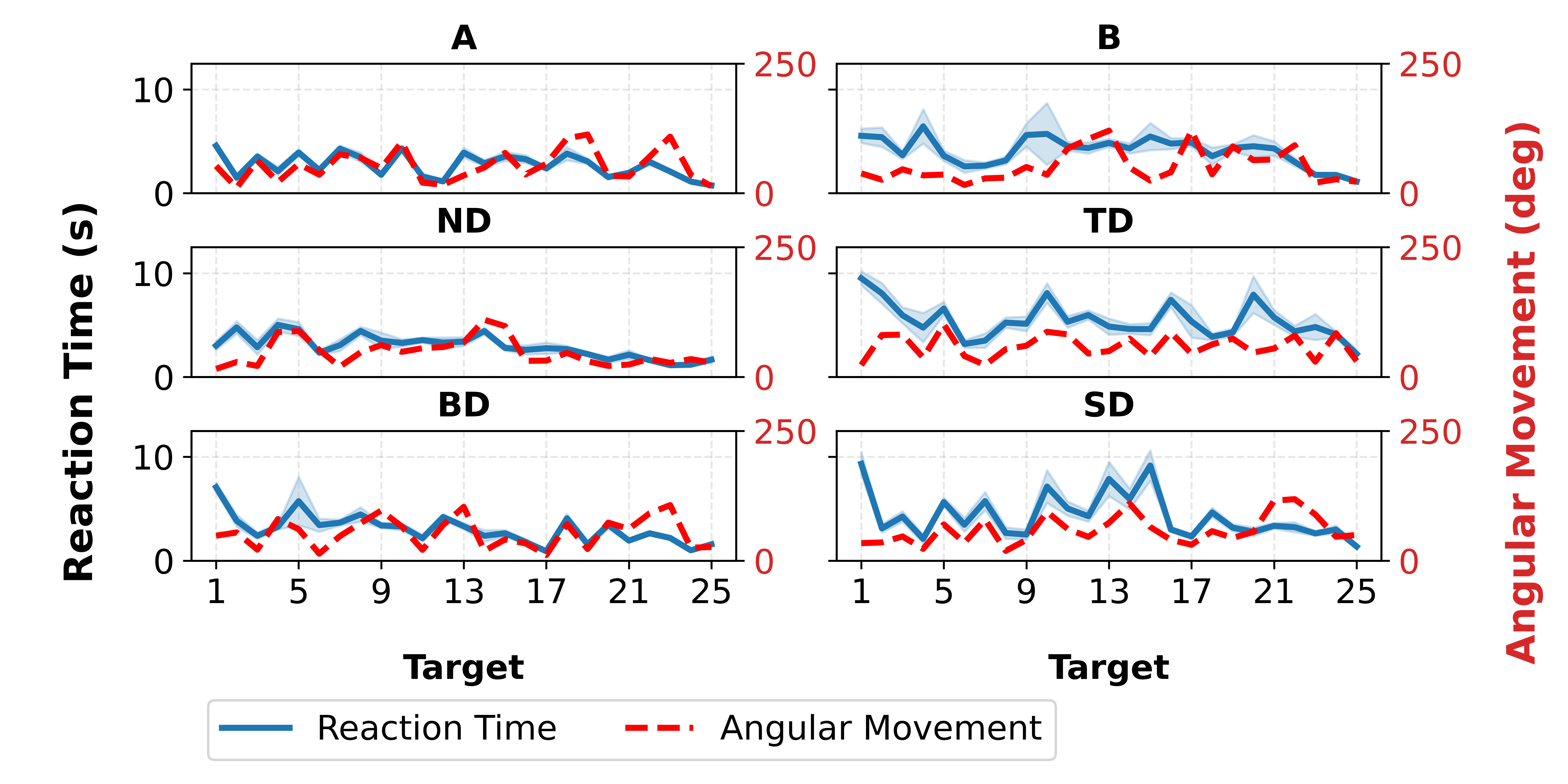}
    \caption{Reaction time (blue) per target and angular movement (red) across six stages. Spatial layout was consistent, with mean per-target angular movement of 53.78° (SD = 4.15°, CV = 7.71\%). Red dotted lines indicate mean angular movement per stage.}
    \label{fig:completion_time_trend}
\end{figure}

\paragraph{Reaction time per target}  Prior work suggests that spatial layout can subtly influence performance in the Trail Making Test~\cite{arnett1995effect, linari2022unveiling}. To examine the relationship between Reaction Time (RT) and spatial demand, we fit an LMM including the interaction between stage and angular movement, and conducted stage-wise LMM analyses to better capture condition-specific effects. Additionally, we examined early-stage distraction effects on first-target RT.
% overal LMM

Fig.~\ref{fig:completion_time_trend} shows target-level RT and required angular head movement across AR-TMT stages. The plot shows average RTs (blue with standard deviation) and angular movement (red dashed) between consecutive targets in each stage. For consistency, angular values were scaled to match the RT range in stage A. 
A positive association was found in stage A ($\beta = 0.022$, \rev{95\% CI [0.013, 0.031]}, $p < .001$), but interaction terms showed no significant differences across stages (all $p$ > .11). 
% Stagewise LMM
Stage-wise LMMs with Holm correction showed that the strongest spatial dependency emerged in ND ($\beta = 0.023$, $p < .001$), followed by stage A ($\beta = 0.022$, $p < .001$), though ND showed an elevated intercept (1.88 s), suggesting higher processing demands due to perceptual clutter. TD ($\beta = 0.019$, $p < .05$), SD ($\beta = 0.018$, $p < .01$), and BD ($\beta = 0.017$, $p < .001$) showed moderate effects, while stage B showed the weakest spatial influence ($\beta = 0.012$, $p < .05$), reflecting greater cognitive switching demands. Overall, spatial–motor demand consistently influenced RT, but weaker correlations in the distraction stages and in stage B compared to A and ND indicate that other cognitive functions contribute more strongly to performance.

\paragraph{First target RT} To assess early-stage distraction effects, we computed angle-adjusted reaction times, as RT was correlated with angular movement. Residual RT was defined as \(\text{Residual RT} = \text{Observed RT} - \alpha \times \text{Angle}\), with \(\alpha = 0.05\,\text{s/deg}\) based on visual axis scaling. An LMM was then fit with Stage as a fixed effect to analyze differences in adjusted RT.

The model revealed longer adjusted RT in TD (\(\beta = 6.42\), \rev{95\% CI [4.86, 7.99]}, \(p < .001\)) and SD (\(\beta = 5.69\), \rev{95\% CI [4.14, 7.26]}, \(p < .001\)) compared to stage A. BD also showed a smaller increase in RT (\(\beta = 2.80\), \rev{95\% CI [1.16, 4.29]}, \(p = .002\)). No significant differences were found for ND and stage B. Holm-corrected pairwise comparisons of estimated marginal means confirmed that TD was slower than all other stages (\rev{$\Delta = 4.6$--$11.5$~s}, $p < .001$). SD was slower than A, B, and ND (\rev{$\Delta = 3.3$--$4.9$~s}, $p < .02$) but not different from TD. BD was slower than A (\rev{$\Delta = +5.3$}, $p < .001$) and ND (\rev{$\Delta = +4.6$}, $p < .001$) yet faster than TD (\rev{$\Delta = -4.6$}, $p < .001$). ND did not differ from A but was faster than SD and TD (\rev{$\Delta = -4.8$ to $-8.2$}, $p < .001$). This suggests that BD can impact early-stage performance, though the effect was weaker than TD and SD.

\begin{figure}[tbp]
    \centering
    \includegraphics[width=\linewidth ,  trim=0 0 0 0, clip]{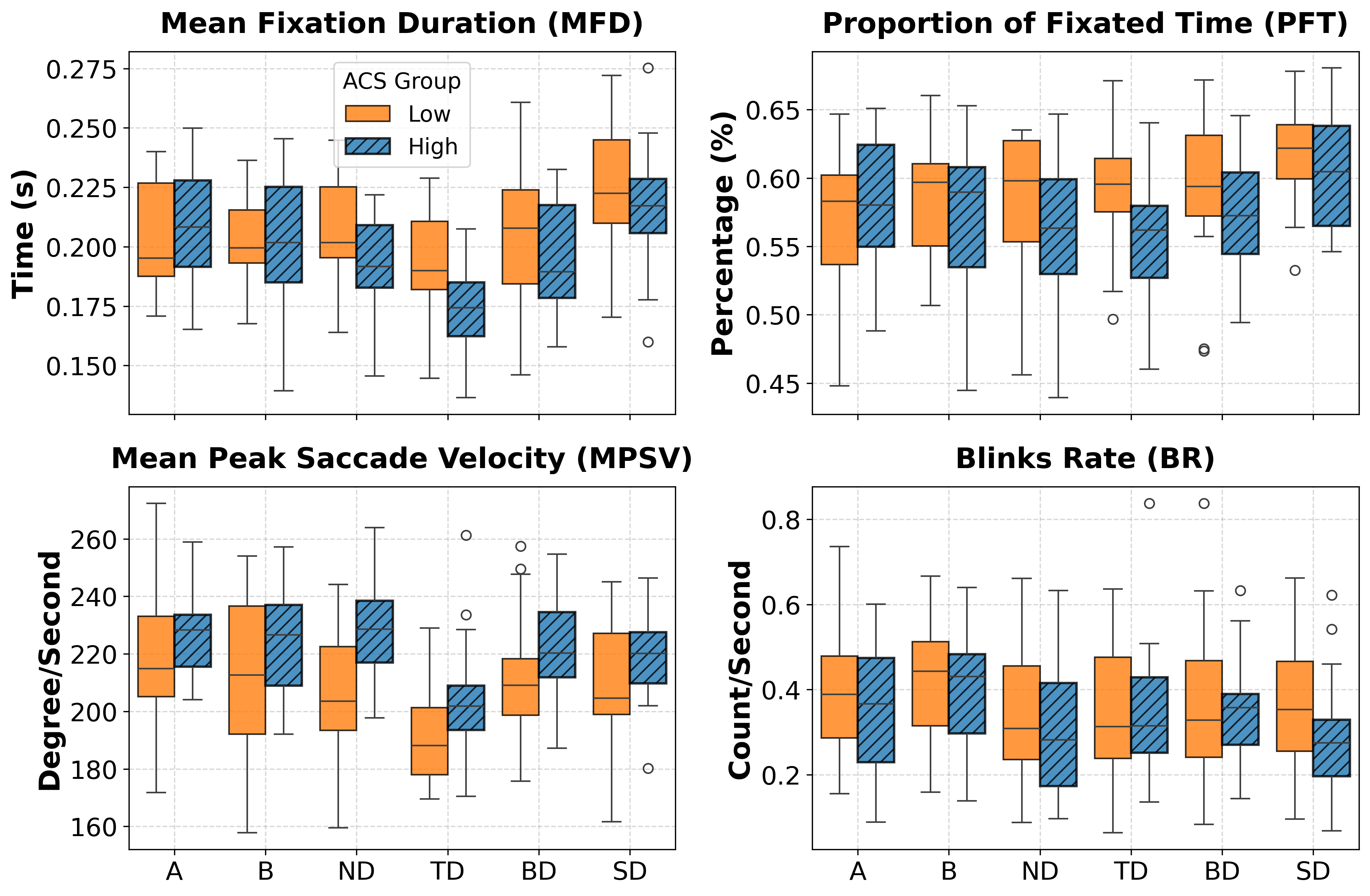}
    \caption{Boxplots of four visual behavior metrics across AR-TMT stages and ACS groups (high vs. low). Only metrics with at least one significant interaction with ACS in the LMM are displayed. Outliers were excluded using the 1.5× interquartile range rule for clarity.}
    \label{fig:visual_behavior}
\end{figure}

\subsection{Visual Behavior Analysis}

This section examines how attention control shapes visual behavior. We analyzed gaze behavior using LMMs with Stage and ACS group (High vs. Low; median split at 37.5\footnote{A prior validation study of the Flanker Squared Test reported an average score of about 39, but we used a median split (37.5) to align group classification with our sample’s distribution.}) as fixed effects to capture potential threshold or nonlinear effects of attention control~\cite{maccallum2002practice, decoster2011best}. Exploratory analyses using ACS as a continuous predictor yielded limited significant effects for gaze and motor metrics, suggesting that attention control may manifest more strongly in overall task outcomes than in fine-grained gaze or motor measures.

\paragraph{Fixation Metrics.}
% MFD , PFT, FR 
% MFD
\textit{Mean Fixation Duration (MFD)} decreased in BD ($\beta = -0.012$, \rev{95\% CI [-.023, -.001]}, $p = .039$), ND ($\beta = -0.017$, \rev{95\% CI [-.028, -.006]}, $p = .003$), and TD ($\beta = -0.035$, \rev{95\% CI [-.046, -.023]}, $p < .001$) compared to stage A, suggesting higher perceptual load due to increased visual clutter and the presence of competing distractors that demanded more frequent attentional shifts~\cite{liu2022assessing}. 
% PFT
\textit{Proportion of Fixated Time (PFT)} decreased during ND ($\beta = -0.027$, \rev{95\% CI [-.052, -.003]}, $p = .030$) and TD ($\beta = -0.032$, \rev{95\% CI [-.057, -.008]}, $p = .010$) compared to stage A. This reduction may reflect increased visual search demands under distraction, prompting broader gaze distribution and reduced fixation on individual targets. 
% FR
\textit{Fixation Rate (FR)} increased during TD ($\beta = 0.326$, \rev{95\% CI [.221, .430]}, $p < .001$), indicating more frequent visual sampling due to greater target–distractor ambiguity. 

% ACS 
No main effect of ACS group was found for any of the metrics.
% ACS MFD
Interaction effects revealed that participants with lower ACS exhibited longer MFD in ND ($\beta = 0.019$, \rev{95\% CI [.003, .035]}, $p = .020$), SD ($\beta = 0.017$, \rev{95\% CI [.001, .033]}, $p = .033$), and TD ($\beta = 0.020$, \rev{95\% CI [.004, .036]}, $p = .013$), suggesting higher susceptibility to distraction as shown in Fig.~\ref{fig:visual_behavior}. 
% ACS part PFT
Additionally, participants with lower ACS showed higher PFT in ND and TD (both $\beta = 0.056$, \rev{95\% CI [.021, .090]}, $p = .001$), suggesting that participants with lower ACS maintained fixations longer, possibly reflecting less efficient visual processing. Overall, BD, ND, and TD increased visual sampling and shortened fixations, and whereas individuals with lower ACS exhibit longer sustained fixations in ND, SD, and TD, indicating greater difficulty filtering irrelevant stimuli~\cite{zagermann2016measuring}.

\paragraph{Saccade Metrics}
% MSA, MSV, MPSV
\textit{Mean Saccade Amplitude (MSA)} and \textit{Mean Saccade Velocity (MSV)} both decreased during TD ($\beta_\mathrm{MSA} = -1.69$, \rev{95\% CI [-2.26, -1.12]}, $p < .001$; $\beta_\mathrm{MSV} = -10.58$, \rev{95\% CI [-13.96, -7.12]}, $p < .001$), suggesting that TD reduced both the extent and speed of gaze shifts, indicating increased local inspection due to target–distractor similarity. 
% MPSV 
\textit{Mean Peak Saccade Velocity (MPSV)} also decreased during SD ($\beta = -9.38$, \rev{95\% CI [-15.67, -3.10]}, $p = .003$) and TD ($\beta = -21.51$, \rev{95\% CI [-27.72, -15.22]}, $p < .001$), reflecting slower saccades under spatial and top-down distractions.  
% ACS overall
Although no main effect of ACS group was observed, an interaction indicated that participants with lower ACS exhibited slower MPSV in ND ($\beta = -13.17$, \rev{95\% CI [-22.06, -4.28]}, $p = .004$) as shown in Fig.~\ref{fig:visual_behavior}.
% Conclusion
These results suggest that TD and SD reduce both the speed and extent of gaze shifts, while individuals with lower ACS show additional vulnerability even under ND.

\paragraph{Oculomotor Metrics} \textit{Blink Rate (BR)} increased during stage B ($\beta = .052$, \rev{95\% CI [.007, .097]}, $p = .022$) as shown in Fig.~\ref{fig:visual_behavior}, potentially reflecting heightened internal processing demands associated with set-shifting~\cite{zhang2015dopamine,muller2007dopamine}. 
% VD
In contrast, \textit{variance of pupil diameter (VD)} showed no significant stage effects. 
% ACS
While no main effect of ACS group was found for BR and VD, participants with lower ACS exhibited reduced BR in BD ($\beta = -0.090$, \rev{95\% CI [-.153, -.026]}, $p = .005$) and TD ($\beta = -0.079$, \rev{95\% CI [-.142, -.016]}, $p = .014$), consistent with greater cognitive load, which suppresses blink rate~\cite{ledger2013effect}. Together, these findings suggest that BR may be a more reliable indicator of distraction and attention control differences than VD.

\paragraph{Entropy Metrics} To quantify visual scanning efficiency~\cite{shiferaw2019review}, we computed \textit{Spatial Entropy (SE)}, \textit{Transition Entropy (TE)}, and \textit{Temporal Duration Entropy (TDE)} from 3D fixation data. SE captures how widely gaze fixations are dispersed in space, TE reflects the unpredictability of gaze transitions between spatial bins, and TDE captures the temporal variability of fixation durations. 

% SE
\textit{SE} increased in SD ($\beta = 0.269$, \rev{95\% CI [.146, .393]}, $p < .001$) and TD ($\beta = 0.199$, \rev{95\% CI [.076, .322]}, $p = .002$), but decreased in BD ($\beta = -0.195$, \rev{95\% CI [-.318, -.072]}, $p = .002$), indicating broader gaze distribution under SD and TD. The reduced dispersion in BD was unexpected and may reflect peripheral processing of salient distractors, enabling foveal attention to remain on targets. This aligns with the later phase of the signal suppression hypothesis~\cite{stilwell2023role} which posits that even highly salient distractors can be suppressed over time, reducing attentional capture and promoting spatially focused gaze. 
% TE
\textit{TE} increased only in SD ($\beta = 0.217$, \rev{95\% CI [.068, .366]}, $p = .005$), indicating greater gaze unpredictability under spatial distraction.
% TDE
\textit{TDE} decreased in SD ($\beta = -0.246$, \rev{95\% CI [-.449, -.044]}, $p = .017$) and TD ($\beta = -0.311$, \rev{95\% CI [-.514, -.109]}, $p = .003$), suggesting more uniform fixation durations. This may reflect a compensatory strategy under increased load, where participants adopt fixed-paced scanning to cope with reduced guidance. 
% ACS
No significant main or interaction effects of ACS group were observed across metrics.
% conclusion
These findings suggest that SD and TD increase gaze dispersion and unpredictability while reducing temporal variability, indicating more dispersed yet temporally rigid scanning behavior. In contrast, BD led to reduced SE, reflecting more focused gaze patterns due to peripheral processing of salient stimuli.

\paragraph{Fixations on ROI} We calculated \textit{Fixation Count} (FC) on targets and distractors, defined as regions of interest (ROIs), by detecting intersections between the 3D gaze vector and the object direction within an angular threshold (2.5$\times$ diameter for SD, 2.0$\times$ otherwise).
The results showed that FC on targets increased in B ($\beta = 70.71$, \rev{95\% CI [41.96, 99.45]}, $p < .001$), SD ($\beta = 58.77$, \rev{95\% CI [30.02, 87.51]}, $p < .001$), and TD ($\beta = 44.65$, \rev{95\% CI [15.90, 73.39]}, $p = .003$), primarily reflecting prolonged task duration. 
FC on distractors was higher in TD compared to ND ($\beta = 135.71$, \rev{95\% CI [116.04, 155.37]}, $p < .001$). No main effects of ACS group were observed for FC on either targets or distractors. Taken together, these findings suggest that elevated distractor fixations occur only when distractors share semantic similarity with targets, whereas fixations on targets primarily reflect overall task duration.

\subsection{Motor Behavior Analysis}

We examined controller-based motor behavior with two measures: Controller Speed (CS) and Controller Movement Entropy (CME). CS reflects movement speed, computed as average controller path length per second~\cite{moore2020extracting}. CME captures directional irregularity, computed as the entropy of framewise changes in movement direction, with higher values indicating more erratic trajectories~\cite{reinhardt2019entropy, loraas2023distinct}. As in the visual behavior analysis, we fit an LMM, which showed decreases in CS during TD ($\beta = -35.59$, $p < .001$, \rev{95\% CI [-49.83, -21.34]}), SD ($\beta = -25.69$, $p < .001$, \rev{95\% CI [-39.94, -11.45]}), and B ($\beta = -23.55$, $p < .001$, \rev{95\% CI [-37.80, -9.31]}). Similarly, CME decreased in TD ($\beta = -.068$, $p = .013$, \rev{95\% CI [-0.122, -0.014]}), SD ($\beta = -.071$, $p = .010$, \rev{95\% CI [-0.125, -0.017]}), and B ($\beta = -.069$, $p = .013$, \rev{95\% CI [-0.123, -0.015]}) relative to stage A, indicating that cognitively demanding stages (TD, SD, B) led to slower and more constrained controller movement in AR. \rev{These results may reflect extended search intervals, leaving the controller inactive between targets and thereby lowering both measures. In addition, increased task complexity can lead to more constrained, less variable motor behavior~\cite{hong2008entropy}, either due to rigidity induced by attentional overload impairing motor flexibility~\cite{hemond2010distraction} or a deliberate strategy to preserve accuracy under high cognitive load~\cite{heitz2014speed}. Confirming the mechanism requires further analyses in future work.}

\begin{figure}[tbp]
    \centering
    \includegraphics[width=\linewidth,  trim=0 0 0 0, clip]{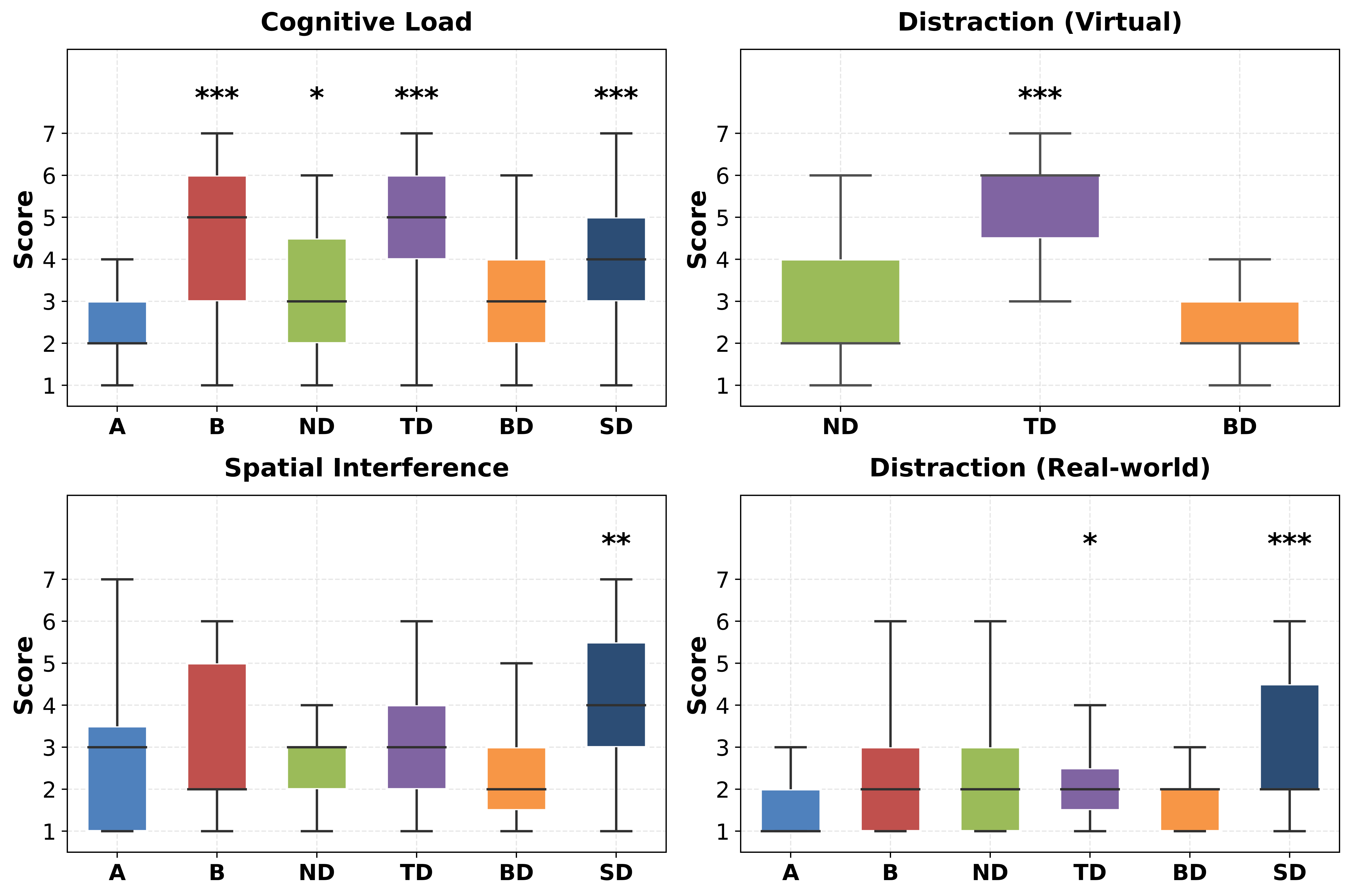}
    \caption{Boxplots of subjective ratings across stages for four question categories: (1) cognitive load, (2) distraction by virtual distractors, (3) comfort with virtual object placement, and (4) distraction by the real-world context.}
    \label{fig:subjective_measures_combined}
\end{figure}

\subsection{Subjective Measures Analysis}

Figure~\ref{fig:subjective_measures_combined} shows boxplots of subjective ratings across stages for four survey measures. To assess the effects of distraction condition and attention control on subjective experience, we fit LMMs for each rating, following the same approach as in the visual behavior analysis.
% Cognitive Load
For \textit{cognitive load}, ratings were higher in TD ($\beta = 2.61$, \rev{95\% CI [1.95, 3.28]}, $p < .001$), B ($\beta = 1.83$, \rev{95\% CI [1.17, 2.50]}, $p < .001$), SD ($\beta = 1.50$, \rev{95\% CI [0.84, 2.16]}, $p < .001$) and ND ($\beta = 0.78$, \rev{95\% CI [0.11, 1.44]}, $p = .022$) compared to stage A, while BD did not differ significantly.
% Distraction from virtual stimuli
For \textit{distraction from virtual stimuli}, we considered only the stages that introduced additional virtual objects (ND, BD, TD). 
Ratings were higher in TD ($\beta = 2.83$, \rev{95\% CI [2.12, 3.54]}, $p < .001$) compared to ND, indicating stronger perceived distraction from top-down distractors.
% Spatial Interfence
For \textit{spatial interference}, only SD showed a significant increase over A ($\beta = 1.39$, \rev{95\% CI [0.54, 2.23]}, $p = .001$), suggesting that loss of spatial anchoring notably impaired perceived spatial stability.
% Real-world distraction 
For \textit{real-world distraction}, ratings were higher in SD 
($\beta = 1.06$, \rev{95\% CI [.43, 1.68]}, $p = .001$) and TD 
($\beta = .67$, \rev{95\% CI [.04, 1.29]}, $p = .037$) compared to stage A, indicating that these stages elicited greater perceived real-world interference. 

% ACS
No significant effects were found for ACS group or its interaction across all measures, indicating a disconnect between perceived distraction and actual attentional performance. This aligns with prior work demonstrating a discrepancy between objective and subjective evaluations of cognitive ability~\cite{harris2023discrepancy}, \rev{suggesting subjective self-reports may not be reliable indicators of actual attention control in AR tasks.} \rev{The stable ratings across some stages may reflect task design features such as clear goals, immediate feedback, and a consistent AR setup, or a ceiling effect, with responses clustering near the upper end of the response scale and reducing sensitivity to subtle differences.}
% Interpretation
Together, distraction stages, especially TD and SD, increased perceived cognitive load and environmental interference, aligning with observed performance drops and gaze changes. 

\section{Discussion}
\label{sec:discussion}

% Cognitive Load

\paragraph{Impacts of Distraction Types} Our findings extend prior research on attention and distraction by demonstrating how distinct types of visual interference impact performance and behavioral patterns in AR environments. Our results show that distraction types grounded in attention-guiding factors engage attention differently. 
% TD
TD elicited substantial performance degradation through semantic similarity, consistent with semantic interference~\cite{burki2020did} or the Stroop effect~\cite{macleod2016stroop}. This was evidenced by longer completion times, delayed responses, higher subjective cognitive load, higher perceptual load reflected in decreased PFT and increased FC on distractors \rev{consistent with H1}, highlighting the challenge of resolving semantic conflict in task-relevant contexts. 
% BD
\rev{As expected in H2}, BD primarily captured attention at onset, as shown by increased first-target reaction times, but its effect was smaller than TD and SD and diminished quickly, with similar completion time to baseline. Unexpectedly, it reduced spatial entropy of gaze, indicating more focused search and effective suppression of distractors once recognized as irrelevant. This aligns with the signal suppression hypothesis, which holds that salient stimuli can be inhibited through top-down control~\cite{stilwell2023role}. 
% SD
\rev{In line with H3,} SD disrupted search organization, with higher spatial and transition entropy indicating less organized gaze patterns, underscoring the importance of contextual and spatial cues for guiding attention in AR environments~\cite{wolfe2021guided}.
Together, these results show that distraction types exert different cognitive impacts in AR, supporting the need to analyze distraction as a set of attention-guiding factors than a single construct.

\paragraph{Impact of Attention Control} Attention control remains an active topic of investigation in psychology, and has only recently begun to receive attention in AR research~\cite{qu2024looking}. \rev{In line with H4}, our findings show that individual differences in attention control are associated with performance during BD and TD, as well as in ND. Prior work shows that attention control is most engaged when distractors match the attentional-control settings defined by task demands~\cite{folk1992involuntary}, and most taxed when distractors resemble the mental representations prioritized in the task set~\cite{oberauer2024meaning}. Therefore, distractors resembling task-set representations increase attentional capture and require greater executive control, making attention control a predictor of performance. In contrast, no significant correlations with ACS were observed for baseline and SD, likely because baseline tasks lack sensitivity to attention control, and SD relies more on spatial representations and working memory than on interference suppression~\cite{szczepanski2010mechanisms, awh2001overlapping}.

\paragraph{Implications} 
\rev{
% Building sentence
Our benchmarking analysis showed that \mbox{TMT-B} in AR correlated with 2D TMT, whereas TMT-A in AR correlated with AR motor speed. This suggests that AR-TMT engages AR-specific aspects, such as 3D visuomotor integration and spatial coordination, offering opportunities to investigate principles of distraction in AR.
% General principles
AR-TMT shows that distractions can be categorized through various factors of attentional guidance, each with different impacts across a range of measures in AR environments.
% Impliaction
Our findings have the potential to generalize to ecologically relevant AR tasks via these principles.
% example
For example, cognitive security studies in AR~\cite{teymourian2025sok, wang2024dark} could investigate the underlying cognitive mechanisms of adversarial attacks such as salience capture in sensory overload or disrupted spatial anchoring in movement manipulation and their potential effects on users.
% ex2 
Additionally, AR interface design could reduce top-down interference by avoiding elements semantically similar to the task goal, such as icons resembling target components in assembly tasks or labels echoing destination names in navigation tasks. Interfaces could also dynamically adjust distraction management strategies based on individual attention control, such as simplifying visual clutter for users with lower attention control or enabling complex overlays for those with higher attention control.
}

\section{Limitations and Future Work} 
\label{sec:limF}

\rev{Attentional performance can vary with temporal factors such as mental fatigue~\cite{boksem2005effects} and circadian rhythms~\cite{valdez2019circadian}. Although we enforced 60\,s inter-stage breaks, we assessed attention control only once and did not record the session time of day or momentary fatigue. In future work, we will improve trait-level characterization by administering repeated attention control assessments across sessions, measuring biomarkers of fatigue (e.g., heart rate variability or salivary cortisol)~\cite{kunasegaran2023understanding}, and using actigraphy to index circadian rhythms~\cite{anna2007measure} to improve trait estimates.} 

\rev{Additionally, we will extend this work by broadening the participant population and the set of distraction types. Our sample size (N=34) was modest and consisted mostly of young, AR-novice adults, which may limit generalization or underrepresent the population distribution of the attention-control trait. Additionally, we did not examine other factors of attentional guidance such as reward or search history~\cite{wolfe2017five, wolfe2021guided}. In future work, we will recruit a larger, demographically broader cohort by engaging XR-expert communities to capture a more experienced population, and integrate additional manipulations (e.g., monetary distraction or adversarial priming) to examine their effects.}

\section{Conclusion}
\label{sec:conclusion}
We present AR-TMT, a framework to comprehensively evaluate how different types of distraction, characterized by the Guided Search model, affect user attention and behavior across multiple measures. Our results show that distraction types shape user behavior in different ways depending on attention-guiding factors, and attention control affects vulnerability under object-based distractions, underscoring the role of environmental context and individual cognitive traits in shaping distraction. These findings have the potential to generalize to ecologically relevant AR tasks, thereby supporting the design of AR systems that address distraction in AR environments.

\begin{acks}
We thank the participants of our user study for their invaluable help in this research. This work was supported in part by NSF grants CSR-2312760, CNS-2112562, and IIS-2231975, NSF CAREER Award IIS-2046072, NSF NAIAD Award 2332744, a Cisco Research Award, a Meta Research Award, Defense Advanced Research Projects Agency Young Faculty Award HR0011-24-1-0001, and the Army Research Laboratory under Cooperative Agreement Number W911NF-23-2-0224. The views and conclusions contained in this document are those of the authors and should not be interpreted as representing the official policies, either expressed or implied, of the Defense Advanced Research Projects Agency, the Army Research Laboratory, or the U.S. Government. This paper has been approved for public release; distribution is unlimited. No official endorsement should be inferred. The U.S.~Government is authorized to reproduce and distribute reprints for Government purposes notwithstanding any copyright notation herein.
\end{acks}

\bibliographystyle{ACM-Reference-Format}
\bibliography{reference}

\end{document}